\documentclass[natbib]{svjour3}

\usepackage{hyperref}
\usepackage{graphicx}
\usepackage{amssymb}

\begin{document}

\title{Influence of fast interstellar gas flow on the dynamics of dust grains}


\author{P. P\'{a}stor}

\institute{P. P\'{a}stor \at
      Tekov Observatory, \\
      Sokolovsk\'{a} 21, 934~01, Levice, Slovak Republic \\
      \email{pavol.pastor@hvezdarenlevice.sk}
}
\date{}

\maketitle

\begin{abstract}
The orbital evolution of a dust particle under the action of a fast
interstellar gas flow is investigated. The secular time derivatives of
Keplerian orbital elements and the radial, transversal, and normal
components of the gas flow velocity vector at the pericentre of
the particle's orbit are derived. The secular time derivatives
of the semi-major axis, eccentricity, and of the radial, transversal,
and normal components of the gas flow velocity vector at the pericentre of
the particle's orbit constitute a system of equations that determines
the evolution of the particle's orbit in space with respect to the gas
flow velocity vector. This system of differential equations
can be easily solved analytically. From the solution of the system
we found the evolution of the Keplerian orbital elements in the special case
when the orbital elements are determined with respect to a plane
perpendicular to the gas flow velocity vector.
Transformation of the Keplerian orbital elements determined
for this special case into orbital elements determined with respect
to an arbitrary oriented plane is presented. The orbital elements
of the dust particle change periodically with a constant oscillation
period or remain constant. Planar, perpendicular and stationary
solutions are discussed.

The applicability of this solution in the Solar system is also investigated.
We consider icy particles with radii from 1 to 10 $\mu$m. The presented
solution is valid for these particles in orbits with semi-major axes from
200 to 3000 AU and eccentricities smaller than 0.8, approximately.
The oscillation periods for these orbits range from 10$^{5}$ to
2 $\times$ 10$^{6}$ years, approximately.

\end{abstract}

\section{Introduction}
\label{sec:introduction}

Since \cite{poynting} and \cite{robertson}, the dynamics of dust
grains has been investigated in many papers and its analysis
is an inseparable part of astrophysics. The influence
of the electromagnetic radiation of a central star is usually taken
into account in the form of the Poynting-Robertson (P-R) effect
(\citealt{poynting,robertson,wywh,burns,klacka}). Beside the P-R effect,
the stellar wind (corpuscular radiation of the star) can also affect
the motion of dust particles. Covariant derivation of the acceleration
caused by the stellar wind and its effects on the dynamics
of dust grains were presented in \cite{SW}.
The Solar magnetic field can affect the motion of charged dust particles
in the Solar system (\citealt{parker,kimura}). Planets can
capture the dust particles into the mean-motion resonances
(\citealt{dermott,reach}). Because of the relative
motion of the Sun with respect to the local interstellar
medium, atoms of the interstellar medium approach the Sun.
The direction of the approach is given by the actual velocity
of the Sun with respect to the local interstellar
medium. These approaching atoms form an interstellar gas flow.
This flow of interstellar atoms through the Solar system has been
already investigated in the past (e.g. \citealt{fahr,mobius,alouani}).
This interstellar gas flow can affect the dynamics
of dust grains in outer parts of the Solar system. The assumption
that dust particles in orbits around other stars can be also affected
by interstellar gas flow is supported by the recent detection
of debris disks around stars with asymmetric morphology caused by
a fast motion of the disk through a cloud of interstellar matter
(\citealt{hines,debes}).

The influence of interstellar gas flow on the dynamics of
a spherical dust particle was investigated by \cite{scherer},
who calculated the secular time derivatives of the particle's
angular momentum and the Laplace-Runge-Lenz vector caused by
the interstellar gas flow. He has come to the conclusion that the particle's
semi-major axis can increase exponentially (\citealt[p. 334]{scherer}).
This result contradicts results of \cite{flow}.
Here the secular time derivatives of the Keplerian orbital
elements of the dust particle under the action of a fast interstellar gas flow
were for the first time calculated for arbitrary orientations of the orbit.
\cite{flow} states that the secular semi-major
axis of the dust particle must decrease under the action of the interstellar
gas flow. The rate of decrease is proportional to the semi-major axis.
Decrease of the semi-major axis only happens in secular
time-scales. Therefore, the exact calculation of the secular
time derivative of the semi-major axis was necessary. This result
was confirmed also by \cite{bera} who investigated
the motion of a dust particle in the outer region of the Solar system
behind the solar wind termination shock. They calculated the orbital
evolution of the dust particle under the action of a constant
mono-directional force, i.e., they solved a case of the classical
Stark problem (for more information about the Stark problem see
\citealt{lantoine} and the references therein). In
the Stark problem it is assumed that the orbital
speed of the dust grain with respect to a central
object can be neglected in comparison with the speed of the interstellar
gas flow. The relative velocity terms are in \cite{flow} taken
into account to the first order of accuracy in the calculation of the secular
time derivatives of Keplerian orbital elements. \cite{bera}
reproduced the result of \cite{flow} on the secular evolution
of the semi-major axis of the particle's orbit. However, the case of
a mono-directional force (i.e., the Stark problem) can be
important for those stars where the relative speed of the neutral gas with
respect to the central star is high. Such a situation can occur, for example,
in merging galaxies when a star from the first galaxy moves through
a molecular cloud of the second galaxy.

\cite{bera} used the orbit-averaged Hamiltonian method.
In this paper we use a method based on orbit-averaged time derivatives of
Keplerian orbital elements. Using this different method we confirm
results of \cite{bera}. We not only use a new method, we also
find new results. We find an explicit form for the time dependence of
all Keplerian orbital elements determined with respect to a reference
plane perpendicular to the gas flow velocity vector. Mainly the time
dependence of the longitude of the ascending node will represent
a generalization of \cite{bera} results. We find a transformation for
the orbital elements determined with respect to the reference plane
perpendicular to the gas flow velocity vector into an arbitrarily
oriented reference plane. We determine the maximal and minimal values
of eccentricity for numerous special cases. We study in detail
the behaviour of the solution for the planar case and for the case when
the gas flow velocity vector is perpendicular to the line of apsides.
Properties of the solution in the Solar system are discussed.

\section{Equation of motion}
\label{sec:eom}

In order to find the acceleration of a spherical dust particle
caused by an interstellar gas flow we will assume
(a) that the dimensions of the dust particle are small in
comparison with the mean free path of the interstellar
gas atoms,
(b) that the mass of the dust particle is large in
comparison with the mass of the interstellar gas atoms, and
(c) that the molecules are specularly or diffusely reflected from
the surface of the dust particle.
These assumptions lead to the following acceleration
(\citealt{baines})
\begin{equation}\label{floweqom}
\frac{d \vec{v}}{dt} = - ~c_{D} ~\gamma_{H} ~
\vert \vec{v} - \vec{v}_{H} \vert ~\left ( \vec{v} - \vec{v}_{H} \right ) ~,
\end{equation}
where $\vec{v}$ is the velocity of the dust grain in the stationary frame,
$\vec{v}_{H}$ is the constant velocity vector of atoms in the flow
in the stationary frame, $c_{D}$ is the drag coefficient,
and $\gamma_{H}$ is the collision parameter.
For the collision parameter we can write
\begin{equation}\label{colpar}
\gamma_{H} = n_{H} ~\frac{m_{H}}{m} ~A ~,
\end{equation}
where $m_{H}$ is the mass of the neutral hydrogen atom,
$n_{H}$ is the concentration of the interstellar neutral
hydrogen atoms, and $A$ $=$ $\pi {R}^{2}$ is the geometrical
cross section of a spherical dust grain of radius $R$
and mass $m$. Assumption (a) is a reasonable
approximation to conditions in interstellar space. Therefore,
Eq. (\ref{floweqom}) is valid for all dust particles with dimensions for
which assumption (b) holds.

In order to find the final equation of motion we further assume (d)
that the dust particle orbits around a star. Therefore we take into account
also the electromagnetic radiation of the star in the form of the P-R effect.
The equation of motion then has the form
\begin{eqnarray}\label{ceqom}
\frac{d \vec{v}}{dt} &=& - ~\frac{\mu}{r^{2}} ~
      \vec{e}_{R}
\nonumber \\
& &   + ~\beta ~\frac{\mu}{r^{2}}
      \left [\left (1 -
      \frac{\vec{v} \cdot \vec{e}_{R}}{c} \right ) \vec{e}_{R}
      - \frac{\vec{v}}{c} \right ]
\nonumber \\
& &   - ~c_{D} ~\gamma_{H} ~
\vert \vec{v} - \vec{v}_{H} \vert ~\left (\vec{v} - \vec{v}_{H} \right ) ~.
\end{eqnarray}
where $\mu$ $=$ $G M$, $G$ is the gravitational constant,
$M$ is the mass of the star, $\vec{r}$ is the position vector
of the particle with respect to the star, $r$ $=$ $\vert \vec{r} \vert$,
$\vec{e}_{R}$ $=$ $\vec{r} / r$, and $c$ is the speed of light in vacuum.
Parameter $\beta$ is defined as the ratio of the electromagnetic
radiation pressure force and the gravitational force between
the star and the particle at rest with respect to the star
\begin{equation}\label{beta}
\beta = \frac{3 ~L ~\bar{Q}'_{pr}}{16 ~\pi ~c ~\mu ~R ~\varrho} ~.
\end{equation}
Here, $L$ is the star luminosity, $\bar{Q}'_{pr}$ is the dimensionless
efficiency factor for radiation pressure integrated over the star spectrum
and calculated for the radial direction ($\bar{Q}'_{pr}$ $=$ 1 for
a perfectly absorbing sphere), and $\varrho$ is the mass density of
the particle. The second term represents the P-R effect neglecting terms
of higher orders in $\vec{v} / c$ than the first (\citealt{klacka}).
Eq. \ref{ceqom} was used in numerical experiments by \cite{marzari}.

We also restrict the possible speeds of the interstellar gas flow. We will
assume (e) that the speed of the interstellar gas flow is much greater
that the mean thermal speed of the gas in the flow. For interstellar
gas flow speeds $v_{H}$ $=$ $\vert \vec{v}_{H} \vert$ comparable to
the mean thermal speed of the gas $c_D$ is a function of $v_{H}$
(\citealt{baines}). However, $c_D$ has an approximately constant value
for those interstellar gas flow speeds much greater that the mean thermal
speed of the gas (\citealt{baines,bana,scherer,NOVA}).

We also assume (f) that the speed of the interstellar gas flow
is much greater than the speed of the dust grain in the stationary
frame associated with the central star.

Finally, we assume (g) that the secular time derivatives of the semi-major
axis and the eccentricity of the particle's orbit caused by the P-R
effect during orbital evolution are low in comparison with
the values of the semi-major axis and the eccentricity, respectively. This
assumption is reasonable for grains with large semi-major axes and
small eccentricities (\citealt{wywh}). These particles have small
orbital speeds. Therefore, in the P-R effect, the terms depending on
velocity can be neglected in comparison with the Keplerian term.

Assumptions (a)--(g) lead to the final equation of motion
\begin{eqnarray}\label{fineqom}
\frac{d \vec{v}}{dt} &=& - ~\frac{\mu ~(1 - \beta)}{r^{2}} ~\vec{e}_{R} +
      c_{D} ~\gamma_{H} ~v_{H} ~\vec{v}_{H}
\nonumber \\
&=&   - ~\frac{\mu_{\beta}}{r^{2}} ~\vec{e}_{R} + \alpha ~\vec{v}_{H} ~.
\end{eqnarray}
If the radial stellar wind is also considered in Eq. (\ref{ceqom}),
then Eq. (\ref{fineqom}) remain unchanged (\citealt{SW}). However,
assumptions (f) and (g) require radial distances that are probably
larger than the radial distance to the stellar wind termination shock for
the majority of Sun-like stars.

\section{Secular evolution of Keplerian orbital elements}
\label{sec:average}

\begin{figure}[t]
\begin{center}
\includegraphics[height=0.15\textheight]{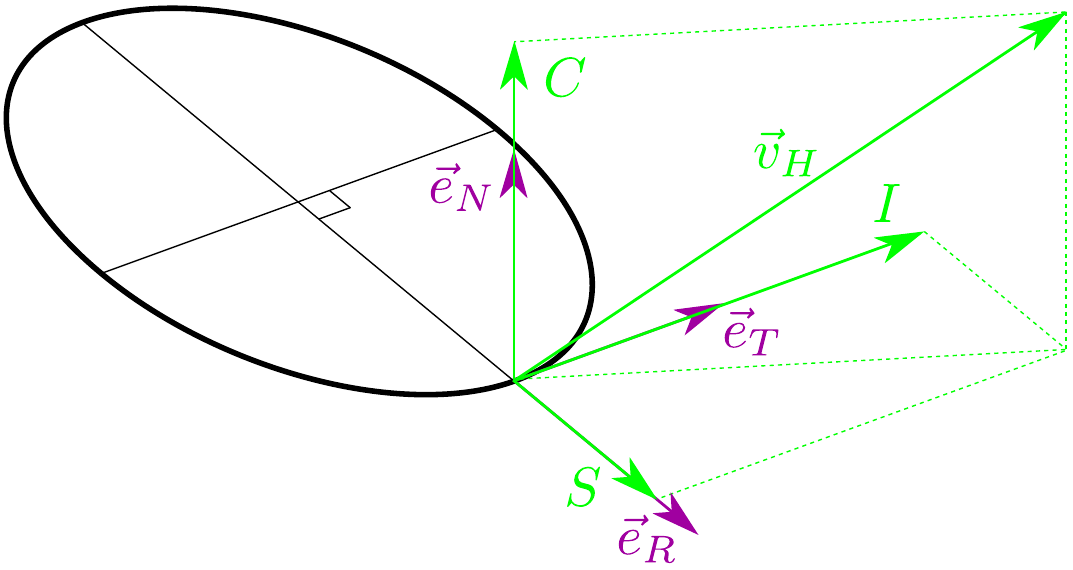}
\end{center}
\caption{A schematic representation of the values $S$, $I$ and $C$ for a given
orbit.}
\label{F1}
\end{figure}

We want to find the influence of the constant acceleration given by
the second term in Eq. (\ref{fineqom}) on the secular evolution
of a particle's orbit. We will assume (h) that constant acceleration
can be considered as a perturbation acceleration to the central
acceleration caused by the gravity of the central star. For the applicability
of the assumptions (f)--(h) in the Solar system we refer the reader
to Appendix \ref{sec:applicability}. We denote the components
of the hydrogen gas velocity vector in the stationary
Cartesian frame associated with the central star as
$\vec{v}_{H}$ $=$ $(v_{HX},v_{HY},v_{HZ})$. In order to
compute the secular time derivatives of the Keplerian orbital
elements ($a$, the semi-major axis; $e$, the eccentricity;
$\omega$, the argument of the pericentre; $\Omega$, the longitude
of the ascending node; $i$, the inclination)
we want to use the Gauss perturbation equations of celestial mechanics.
To do this, we need to know the radial, transversal, and normal
components of acceleration given by the second term in Eq. (\ref{fineqom}).
The orthogonal radial, transversal, and normal unit vectors of the particle
on a Keplerian orbit are (e.g., \citealt{relation})
\begin{eqnarray}
\label{er}
\vec{e}_{R} &=& \left (\cos \Omega ~\cos (f+ \omega) -
      \sin \Omega ~\sin (f+ \omega) ~\cos i ~, \right.
\nonumber \\
& &   \left. \sin \Omega ~\cos (f+ \omega) +
      \cos \Omega ~\sin (f+ \omega) ~\cos i ~, \right.
\nonumber \\
& &   \left. \sin (f+ \omega) ~\sin i \right ) ~,\\
\label{et}
\vec{e}_{T} &=& \left ( - \cos \Omega ~\sin (f+ \omega) -
      \sin \Omega ~\cos (f+ \omega) ~\cos i ~, \right.
\nonumber \\
& &   \left. - \sin \Omega ~\sin (f+ \omega) +
      \cos \Omega ~\cos (f+ \omega) ~\cos i ~, \right.
\nonumber \\
& &   \left. \cos (f+ \omega) ~\sin i \right ) ~,\\
\label{en}
\vec{e}_{N} &=& (\sin \Omega ~\sin i, ~- \cos \Omega ~\sin i, ~\cos i) ~,
\end{eqnarray}
where $f$ is the true anomaly. For the radial, transversal and normal
components of the perturbation acceleration, we obtain
\begin{eqnarray}
\label{ar}
a_{R} &=& \alpha ~\vec{v}_{H} \cdot \vec{e}_{R} = \alpha ~A ~,\\
\label{at}
a_{T} &=& \alpha ~\vec{v}_{H} \cdot \vec{e}_{T} = \alpha ~B ~,\\
\label{an}
a_{N} &=& \alpha ~\vec{v}_{H} \cdot \vec{e}_{N} = \alpha ~C ~.
\end{eqnarray}
Now we can use the Gauss perturbation equations of celestial mechanics
to compute the time derivatives of the orbital elements. The perturbation
equations have the form (cf., e.g., \citealt{mude,danby})
\begin{eqnarray}\label{gauss}
\frac{d a}{d t} &=& \frac{2~a}{1 - e^{2}} ~
      \sqrt{\frac{p}{\mu_{\beta}}} ~
      \left [ a_{R} ~e ~\sin f +
      a_{T} \left ( 1 + e \cos f \right ) \right ] ~,
\nonumber \\
\frac{d e}{d t} &=&
      \sqrt{\frac{p}{\mu_{\beta}}} ~\left [ a_{R} ~\sin f +
      a_{T} \left (\cos f +
      \frac{e + \cos f}{1 + e \cos f} \right ) \right ] ~,
\nonumber \\
\frac{d \omega}{d t} &=& - ~\frac{1}{e} ~\sqrt{\frac{p}{\mu_{\beta}}} ~
      \left ( a_{R} ~\cos f - a_{T} ~
      \frac{2 + e \cos f}{1 + e \cos f} ~
      \sin f \right )
\nonumber \\
& &   - ~\frac{r}{\sqrt{\mu_{\beta} p}} ~
      a_{N} ~\frac{\sin (f + \omega)}{\sin i} ~\cos i ~,
\nonumber \\
\frac{d \Omega}{d t} &=&
      \frac{r}{\sqrt{\mu_{\beta} p}} ~
      a_{N} ~\frac{\sin (f + \omega)}{\sin i} ~,
\nonumber \\
\frac{d i}{d t} &=& \frac{r}{\sqrt{\mu_{\beta} p}} ~a_{N} ~
      \cos (f + \omega) ~,
\end{eqnarray}
where $p$ $=$ $a (1 - e^{2})$. The time average of any quantity $g$ during
one orbital period $T$ can be computed using
\begin{eqnarray}\label{average}
\left \langle g \right \rangle &=& \frac{1}{T} \int_{0}^{T} g ~dt =
      \frac{\sqrt{\mu_{\beta}}}{2 ~\pi ~a^{3/2}} \int_{0}^{2 \pi} g
      \left (\frac{df}{dt} \right )^{-1} df
\nonumber \\
&=&   \frac{\sqrt{\mu_{\beta}}}{2 ~\pi ~a^{3/2}} \int_{0}^{2 \pi} g
      \left (\frac{\sqrt{\mu_{\beta} p}}{r^{2}} \right )^{-1} df
\nonumber \\
&=&   \frac{1}{2 ~\pi ~a^{2} ~\sqrt{1 - e^{2}}}
      \int_{0}^{2 \pi} g ~r^{2} ~df ~,
\end{eqnarray}
where the second ($\sqrt{\mu_{\beta} p}$ $=$ $r^{2} df/dt$) and the third
($4 \pi^{2} a^{3}$ $=$ $\mu_{\beta} T^{2}$) of Kepler's laws were used.
From Eqs. (\ref{ar})--(\ref{average}) we finally obtain for the secular
time derivatives of the Keplerian orbital elements
\begin{eqnarray}
\label{dadt}
\left \langle \frac{da}{dt} \right \rangle &=& 0 ~,\\
\label{dedt}
\left \langle \frac{de}{dt} \right \rangle &=& \frac{3 ~\alpha}{2} ~
      \sqrt{\frac{p}{\mu_{\beta}}} ~I ~,\\
\label{dwdt}
\left \langle \frac{d \omega}{dt} \right \rangle &=& - ~
      \frac{3 ~\alpha}{2} ~\sqrt{\frac{p}{\mu_{\beta}}} ~
      \left (\frac{S}{e} - C ~\frac{\cos i}{\sin i} ~
      \frac{e \sin \omega}{1 - e^{2}} \right ) ~,\\
\label{dOdt}
\left \langle \frac{d \Omega}{dt} \right \rangle &=& - ~
      \frac{3 ~\alpha}{2} ~
      \sqrt{\frac{p}{\mu_{\beta}}} ~\frac{C}{\sin i} ~
      \frac{e ~\sin \omega}{1 - e^{2}} ~,\\
\label{didt}
\left \langle \frac{di}{dt} \right \rangle &=& - ~
      \frac{3 ~\alpha}{2} ~
      \sqrt{\frac{p}{\mu_{\beta}}} ~C ~
      \frac{e ~\cos \omega}{1 - e^{2}} ~,
\end{eqnarray}
where the quantities
\begin{eqnarray}\label{SIC}
S &=& (\cos \Omega ~\cos \omega -
      \sin \Omega ~\sin \omega ~\cos i) ~v_{HX}
\nonumber \\
& &   + ~(\sin \Omega ~\cos \omega +
      \cos \Omega ~\sin \omega ~\cos i) ~v_{HY}
\nonumber \\
& &   + ~\sin \omega ~\sin i ~v_{HZ} ~,
\nonumber \\
I &=& (- \cos \Omega ~\sin \omega -
      \sin \Omega ~\cos \omega ~\cos i) ~v_{HX}
\nonumber \\
& &   + ~(- \sin \Omega ~\sin \omega +
      \cos \Omega ~\cos \omega ~\cos i) ~v_{HY}
\nonumber \\
& &   + ~\cos \omega ~\sin i ~v_{HZ} ~,
\nonumber \\
C &=& \sin \Omega ~\sin i ~v_{HX} - \cos \Omega ~\sin i ~v_{HY} +
      \cos i ~v_{HZ} ~,
\end{eqnarray}
are values of
$A$ $=$ $\vec{v}_{H} \cdot \vec{e}_{R}$,
$B$ $=$ $\vec{v}_{H} \cdot \vec{e}_{T}$ and
$C$ $=$ $\vec{v}_{H} \cdot \vec{e}_{N}$ at the pericentre of the particle
orbit ($f$ $=$ 0), respectively. The value of $C$ is a constant on
a given oscular orbit. The values of $S$, $I$ and $C$ are depicted in
Fig. \ref{F1}.
For the special case $\vec{v}_{H}$ $=$ $(-v_{H},0,0)$ the system
of Eqs. (\ref{dadt})--(\ref{didt}) and
Eqs. (\ref{SIC}) reduces to the system of Eqs. (3)--(7) in \cite{mihe} with
the exception of one term in the secular time derivative of
the longitude of ascending node. \cite{mihe} studied the motion
of a particle around a planet subject to radiation pressure
from a central star. Mathematically, they considered
the Stark problem in a rotating reference frame. The extra term
in the secular time derivative of the longitude of ascending node
is caused by the Coriolis force. The Stark problem
is recovered if the rotation rate is set to zero. Because of
the extra term caused by the Coriolis force their analytical
solution is different and does not correspond to the solution which will
be given in this paper. Their solution of the Stark
problem is not easy to see from their notation in the rotating
reference frame. Moreover, they considered only the case for which
$\vec{v}_{H}$ lies in the plane with $i$ $=$ 0 and is antiparallel
with $x$ axis. Our solution will be more general.

\section{Secular evolution of orbit with respect to gas flow velocity
vector}
\label{sec:SIC}

The parameters $S$, $I$ and $C$ determine the position of the orbit with
respect to the gas flow velocity vector. Therefore, their time derivatives
are useful for a description of the evolution of the orbit's position
in space. Putting Eqs. (\ref{dwdt})--(\ref{didt}) into the formulas for
the averaged time derivatives of the quantities $S$, $I$ and $C$ given by
Eqs. (\ref{SIC}), we obtain
\begin{eqnarray}
\label{dSdt}
\left \langle \frac{dS}{dt} \right \rangle &=& - ~
      \frac{3 ~\alpha}{2} ~\sqrt{\frac{p}{\mu_{\beta}}} ~
      \frac{S ~I}{e} ~,\\
\label{dIdt}
\left \langle \frac{dI}{dt} \right \rangle &=& - ~
      \frac{3 ~\alpha}{2} ~\sqrt{\frac{p}{\mu_{\beta}}} ~
      \left (\frac{eC^{2}}{1 - e^{2}} - \frac{S^{2}}{e} \right ) ~,\\
\label{dCdt}
\left \langle \frac{dC}{dt} \right \rangle &=&
      \frac{3 ~\alpha}{2} ~\sqrt{\frac{p}{\mu_{\beta}}} ~
      \frac{e ~I ~C}{1 - e^{2}} ~.
\end{eqnarray}
Eqs. (\ref{dSdt})--(\ref{dCdt}) are not independent, because
$S$ $\langle dS/dt \rangle$ $+$
$I$ $\langle dI/dt \rangle$ $+$
$C$ $\langle dC/dt \rangle$ $=$ 0
always hold. Eqs. (\ref{dSdt})--(\ref{dCdt})
together with Eqs. (\ref{dadt})--(\ref{dedt})
represent the system of equations that determines the evolution
of the particle's orbit in space with respect to the gas flow
velocity vector. All orbits that are created with rotations of one orbit
around the line aligned with the gas flow velocity vector and going
through the centre of gravity will undergo the same evolution
determined by this system of equations.

Now we find the solution of the system of equations given by
Eqs. (\ref{dadt})--(\ref{dedt}) and Eqs. (\ref{dSdt})--(\ref{dCdt}).
If we combine Eq. (\ref{dedt}) with Eq. (\ref{dSdt}), then we obtain
\begin{equation}\label{S1}
\left \langle \frac{dS}{dt} \right \rangle = -
\left \langle \frac{de}{dt} \right \rangle \frac{S}{e} ~.
\end{equation}
This equation leads, for $S$ $\neq$ 0, to the differential equation
\begin{equation}\label{S2}
\frac{dS}{S} = - ~\frac{de}{e} ~,
\end{equation}
with the solution
\begin{equation}\label{S3}
\vert S \vert = \frac{D}{e} ~.
\end{equation}
Here, $D$ is a constant which can be determined from the initial conditions.
Thus, if the major axis of the orbit is aligned with the direction of
the gas flow velocity vector, then the eccentricity is minimal.
If we combine Eq. (\ref{dedt}) with Eq. (\ref{dCdt}), then we obtain
\begin{equation}\label{C1}
\left \langle \frac{dC}{dt} \right \rangle =
\frac{e ~C}{1 - e^{2}} \left \langle \frac{de}{dt} \right \rangle ~.
\end{equation}
This equation leads, for $C$ $\neq$ 0, to the differential equation
\begin{equation}\label{C2}
\frac{dC}{C} = \frac{e ~de}{1 - e^{2}} ~,
\end{equation}
with the solution
\begin{equation}\label{C3}
\vert C \vert = \frac{F}{\sqrt{1 - e^{2}}} ~,
\end{equation}
where $F$ is an integration constant. Thus, if the magnitude of the normal
component of $\vec{v}_{H}$ measured in the perihelion is maximal, then
the eccentricity is maximal. For $I$ we obtain from
$S^{2}$ $+$ $I^{2}$ $+$ $C^{2}$ $=$ $v^{2}_{H}$
\begin{equation}\label{absI1}
\vert I \vert = \sqrt{v^{2}_{H} - \frac{D^{2}}{e^{2}} -
\frac{F^{2}}{1 - e^{2}}} ~.
\end{equation}
If $S$ $=$ 0, then from Eq. (\ref{dSdt}) we obtain
$\langle dS/dt \rangle$ $=$ 0. Since Eq. (\ref{C3}) holds for
the case $S$ $=$ 0 and $C$ $\neq$ 0, also Eq. (\ref{absI1})
with $D$ $=$ 0 holds. Therefore we put $D$ $=$ 0 in
Eq. (\ref{S3}) for the case $S$ $=$ 0 and $C$ $\neq$ 0.
If $C$ $=$ 0, then from Eq. (\ref{dCdt}) we obtain
$\langle dC/dt \rangle$ $=$ 0. Since Eq. (\ref{S3}) holds
for the case $C$ $=$ 0 and $S$ $\neq$ 0, also Eq. (\ref{absI1})
with $F$ $=$ 0 holds. Therefore we put $F$ $=$ 0 in
Eq. (\ref{C3}) for the case $C$ $=$ 0 and $S$ $\neq$ 0.
If $S$ $=$ 0 and $C$ $=$ 0, then we have $\vert I \vert$
$=$ $v_{H}$. Therefore we put $D$ $=$ 0 and $F$ $=$ 0 in
Eq. (\ref{S3}) and Eq. (\ref{C3}), respectively. We come
to the conclusion that Eqs. (\ref{S3}), (\ref{C3}) and
(\ref{absI1}) always hold during the orbital motion of
the particle. Eq. (\ref{S3}) and Eq. (\ref{C3}) together
with the properties of the system of differential equations
given by Eqs. (\ref{dadt})--(\ref{dedt}) and
Eqs. (\ref{dSdt})--(\ref{dCdt}) imply that $S$ and $C$
can not change sign during the orbital evolution of the particle.
If $D$ $\neq$ 0, then both $S$ and $e$ must be non-zero
during the orbital evolution of the particle. Similarly, if
$F$ $\neq$ 0, then $C$ $\neq$ 0 and $e$ $\neq$ 1 during
the orbital evolution of the particle. For the special cases
$D$ $=$ 0 or $F$ $=$ 0 is necessary to use the properties
of the whole system the differential equations given by
Eqs. (\ref{dadt})--(\ref{dedt}) and Eqs. (\ref{dSdt})--(\ref{dCdt}),
because, as we will see later, in these cases the eccentricity
can be close to 0 or 1, respectively. Therefore we can write
\begin{equation}\label{S}
S = \frac{U}{e} ~,
\end{equation}
\begin{equation}\label{C}
C = \frac{V}{\sqrt{1 - e^{2}}}
\end{equation}
and
\begin{equation}\label{absI2}
\vert I \vert = \sqrt{v^{2}_{H} - \frac{U^{2}}{e^{2}} -
\frac{V^{2}}{1 - e^{2}}} ~,
\end{equation}
where $U$ and $V$ are some constants. We come to the conclusion that also
Eqs. (\ref{S})--(\ref{absI2}) always hold during the orbital motion of
the particle.

The $S$, $\vert I \vert$, and $C$ components of the hydrogen gas velocity
vector at the pericentre of the particle's orbit depend only on the particle's
eccentricity. To find the evolution of the eccentricity, we insert
Eq. (\ref{absI2}) into Eq. (\ref{dedt}). We obtain
\begin{eqnarray}\label{dedtI}
\frac{de}{dt} &=& \pm ~\frac{3 ~\alpha ~v_{H}}{2 ~e} ~
      \sqrt{\frac{a}{\mu_{\beta}}} ~
\nonumber \\
& &   \times ~\sqrt{e^{2}(1 - e^{2}) - \frac{U^{2}}{v^{2}_{H}} ~
      (1 - e^{2}) - \frac{V^{2}}{v^{2}_{H}} ~e^{2}} ~.
\end{eqnarray}
The plus sign is for positive values of $I$ and the minus sign for
negative values of $I$. If $I$ is negative, then the eccentricity decreases.
If $I$ is positive, then the eccentricity increases. In order to integrate this
equation we rewrite the expression in the second square root in the following
form
\begin{equation}\label{quad1}
e^{2}(1 - e^{2}) - \frac{U^{2}}{v^{2}_{H}} ~(1 - e^{2}) -
\frac{V^{2}}{v^{2}_{H}} ~e^{2} =
(e^{2}_{1} - e^{2})(e^{2} - e^{2}_{2}) ~,
\end{equation}
where
\begin{equation}\label{e1}
e^{2}_{1} = \frac{1 + \frac{U^{2}}{v^{2}_{H}} -
\frac{V^{2}}{v^{2}_{H}}}{2} + \sqrt{\left (\frac{1 + \frac{U^{2}}
{v^{2}_{H}} - \frac{V^{2}}{v^{2}_{H}}}{2} \right )^{2} -
\frac{U^{2}}{v^{2}_{H}}} ~,
\end{equation}
\begin{equation}\label{e2}
e^{2}_{2} = \frac{1 + \frac{U^{2}}{v^{2}_{H}} -
\frac{V^{2}}{v^{2}_{H}}}{2} - \sqrt{\left (\frac{1 + \frac{U^{2}}
{v^{2}_{H}} - \frac{V^{2}}{v^{2}_{H}}}{2} \right )^{2} -
\frac{U^{2}}{v^{2}_{H}}} ~.
\end{equation}
It is possible to show (using Eqs. \ref{S}, \ref{C} and \ref{absI2})
that the expression in the square root is always positive or zero,
$e^{2}_{1}$ $\in$ [0, 1] and $e^{2}_{2}$ $\in$ [0, 1].
From Eq. (\ref{dedtI}) we obtain for $I$ $\neq$ 0
\begin{eqnarray}\label{eint}
\int \frac{e ~de}{\sqrt{(e^{2}_{1} - e^{2})(e^{2} - e^{2}_{2})}} &=&
      \arcsin \sqrt{\frac{e^{2} - e^{2}_{2}}{e^{2}_{1} - e^{2}_{2}}} + \varphi
\nonumber \\
&=&   \pm ~\frac{3 ~\alpha ~v_{H}}{2} ~\sqrt{\frac{a}{\mu_{\beta}}} ~t ~,
\end{eqnarray}
where $\varphi$ is an integration constant. Hence
\begin{equation}\label{epm}
e^{2} = \frac{e^{2}_{1} + e^{2}_{2}}{2} - \frac{e^{2}_{1} - e^{2}_{2}}{2} ~
\cos \left (\pm ~3 \alpha v_{H} ~\sqrt{\frac{a}{\mu_{\beta}}} ~t -
2 \varphi \right ) ~.
\end{equation}
From this equation we can see that the eccentricity changes
between its minimal value $e_{2}$ and its maximal value $e_{1}$ for
one solution determined by the choice of the sign ($+$ or $-$).
We denote the time close to the maximum eccentricity $e_{1}$ as $t_{M}$.
From Eq. (\ref{epm}) we obtain for values of $\varphi$ at the time $t_{M}$
\begin{equation}\label{ephip}
2 \varphi_{+} = 3 ~\alpha ~v_{H} ~\sqrt{\frac{a}{\mu_{\beta}}} ~t_{M} -\pi -
2 ~k_{1} ~\pi ~,
\end{equation}
\begin{equation}\label{ephim}
2 \varphi_{-} = - ~3 ~\alpha ~v_{H} ~\sqrt{\frac{a}{\mu_{\beta}}} ~t_{M} -\pi -
2 ~k_{2} ~\pi ~,
\end{equation}
where $k_{1}$ and $k_{2}$ are two integers. In Eq. (\ref{epm}) both
these values lead to the same solution
\begin{equation}\label{e}
e^{2} = \frac{e^{2}_{1} + e^{2}_{2}}{2} + \frac{e^{2}_{1} - e^{2}_{2}}{2} ~
\cos \left (3 \alpha v_{H} ~\sqrt{\frac{a}{\mu_{\beta}}} ~
(t - t_{M}) \right ) ~.
\end{equation}
This solution is in accordance with the result of \cite{bera}
up to the definition of an additive constant for the time. The eccentricity
changes periodically with the oscillation period
\begin{equation}\label{Te}
T_{e} = \frac{2 ~\pi}{3 ~\alpha ~v_{H}} ~\sqrt{\frac{\mu_{\beta}}{a}} ~.
\end{equation}
For a connection between this result (obtained from
Eq. \ref{fineqom}) and the results obtained from the equation of motion
with the relative velocity $\vec{v}$ included in the force
caused by the interstellar gas flow, we refer the reader to
Appendix \ref{sec:connection}. The values of $T_{e}$ for dust particles in
the Solar system are in Table \ref{T1}. For the Solar system we used
$L_{\odot}$ $=$ 3.842 $\times$ 10$^{26}$ W (\citealt{bahcall}),
$c_D$ $=$ 2.6 (\citealt{bana,scherer,flow}),
$n_{H}$ $=$ 0.2 cm$^{-3}$ (\citealt{bera}) and
$v_{H}$ $=$ 26 km s$^{-1}$ (e.g. \citealt{lallement,landgraf})
(see Appendix \ref{sec:applicability}).

\begin{table}[t]
\centering
\begin{tabular}{c c c c}
$R$ $=$ 1 $\mu$m & $R$ $=$ 2 $\mu$m & $R$ $=$ 5 $\mu$m & $R$ $=$ 10 $\mu$m \\
\begin{tabular}{| c | c |}
\hline
$a$ & $T_{e}$ \\

[AU] & $\left [10^{5} \atop \mbox{years} \right ]$ \\
\hline
200 & 2.06 \\
300 & 1.68 \\
400 & 1.46 \\
500 & 1.30 \\
600 & 1.19 \\
700 & 1.10 \\
\hline
\end{tabular}
&
\begin{tabular}{| c | c |}
\hline
$a$ & $T_{e}$ \\

[AU] & $\left [10^{5} \atop \mbox{years} \right ]$ \\
\hline
200 & 5.35 \\
400 & 3.78 \\
600 & 3.09 \\
800 & 2.67 \\
1000 & 2.39 \\
1200 & 2.18 \\
\hline
\end{tabular}
&
\begin{tabular}{| c | c |}
\hline
$a$ & $T_{e}$ \\

[AU] & $\left [10^{5} \atop \mbox{years} \right ]$ \\
\hline
200 & 14.90 \\
600 & 8.60 \\
1000 & 6.66 \\
1400 & 5.63 \\
1800 & 4.97 \\
2200 & 4.49 \\
\hline
\end{tabular}
&
\begin{tabular}{| c | c |}
\hline
$a$ & $T_{e}$ \\

[AU] & $\left [10^{5} \atop \mbox{years} \right ]$ \\
\hline
500 & 19.45 \\
1000 & 13.75 \\
1500 & 11.23 \\
2000 & 9.73 \\
2500 & 8.70 \\
3000 & 7.94 \\
\hline
\end{tabular}
\end{tabular}
\caption{Oscillation periods $T_{e}$ determined for orbits with various
semi-major axes in the Solar system. Dust particles with radius $R$
$\in$ \{1 $\mu$m, 2 $\mu$m, 5 $\mu$m, 10 $\mu$m\}, mass density
$\varrho$ $=$ 1 g cm$^{-3}$ and $\bar{Q}'_{pr}$ $=$ 1 are used.}
\label{T1}
\end{table}

To find the time evolution of $S$, $\vert I \vert$ and $C$, we can put
Eq. (\ref{e}) into Eqs. (\ref{S})--(\ref{absI2}). $S$, $\vert I \vert$
and $C$ also change periodically with period $T_{e}$. Now we
find the evolution of $I$ from Eq. (\ref{absI2}). If we put Eq. (\ref{e}) into
Eq. (\ref{quad1}), then we get
\begin{eqnarray}\label{quad2}
(e^{2}_{1} - e^{2})(e^{2} - e^{2}_{2}) &=&
      \left (\frac{e^{2}_{1} - e^{2}_{2}}{2} \right )^{2}
\nonumber \\
& &   \times ~\sin^{2} \left (3 \alpha v_{H} ~\sqrt{\frac{a}{\mu_{\beta}}} ~
      (t - t_{M}) \right ) ~.
\end{eqnarray}
Hence,
\begin{equation}\label{absI3}
\vert I \vert = \frac{v_{H}}{e ~\sqrt{1 - e^{2}}} ~
\frac{e^{2}_{1} - e^{2}_{2}}{2} ~
\left \vert \sin \left (3 \alpha v_{H} ~\sqrt{\frac{a}{\mu_{\beta}}} ~
(t - t_{M}) \right ) \right \vert ~.
\end{equation}
If $I$ is negative, then the secular eccentricity must decrease
(see Eq. \ref{dedt}). Therefore if we compare the evolutions given
by Eq. (\ref{e}) and Eq. (\ref{absI3}), we come to the conclusion that
\begin{equation}\label{I}
I = - ~\frac{v_{H}}{e ~\sqrt{1 - e^{2}}} ~
\frac{e^{2}_{1} - e^{2}_{2}}{2} ~
\sin \left (3 \alpha v_{H} ~\sqrt{\frac{a}{\mu_{\beta}}} ~
(t - t_{M}) \right ) ~.
\end{equation}
Eq. (\ref{e}) and Eq. (\ref{I}) hold also for the special cases
$U$ $=$ 0 or $V$ $=$ 0.

\subsection{Planar case}
\label{subsec:planar}

$C$ $=$ 0 for the special case when the velocity of the hydrogen gas
$\vec{v}_{H}$ lies in the orbital plane of the particle.
Putting $C$ $=$ 0 ($V$ $=$ 0) in Eqs. (\ref{e1})--(\ref{e2}), one gets
\begin{equation}\label{plance1}
e^{2}_{1} = 1 ~,
\end{equation}
\begin{equation}\label{plance2}
e^{2}_{2} = \frac{U^{2}}{v^{2}_{H}} ~.
\end{equation}
Therefore the minimum eccentricity is $\vert U \vert / v_{H}$ and
the maximum eccentricity is 1. Using Eqs. (\ref{plance1})--(\ref{plance2})
we obtain from Eqs. (\ref{S}), (\ref{e}), and (\ref{I}), for the planar case,
\begin{equation}\label{plancS}
S = \frac{U}{e} ~,
\end{equation}
\begin{equation}\label{plancI}
I = - ~\frac{v_{H}}{e ~\sqrt{1 - e^{2}}} ~
\frac{1 - \frac{U^{2}}{v^{2}_{H}}}{2} ~
\sin \left (3 \alpha v_{H} ~\sqrt{\frac{a}{\mu_{\beta}}} ~
(t - t_{M}) \right ) ~,
\end{equation}
\begin{equation}\label{plance}
e^{2} = 1 - \left (1 - \frac{U^{2}}{v^{2}_{H}} \right )
\sin^{2} \left (\frac{3 \alpha v_{H}}{2} ~\sqrt{\frac{a}{\mu_{\beta}}} ~
(t - t_{M}) \right ) ~.
\end{equation}
We can mention that $S^{2}$ $+$ $I^{2}$ $=$ $v^{2}_{H}$ always holds
for the planar case. Eq. (\ref{plance}) is in accordance with the result from
\cite{bera} for the planar case up to the definition
of an additive constant for the time.

For the planar case we obtain from Eq. (\ref{dIdt})
\begin{equation}\label{plancdIdt}
\left \langle \frac{dI}{dt} \right \rangle = \frac{3 ~\alpha}{2} ~
\sqrt{\frac{p}{\mu_{\beta}}} ~\frac{S^{2}}{e} \geq 0 ~.
\end{equation}
Therefore, in the planar case $I$ increases with time. This is
in accordance with Eq. (\ref{plancI}). To show this we rewrite
Eq. (\ref{plancI}) as
\begin{eqnarray}\label{plancIchange}
I &=& - ~\frac{v_{H}}{e} ~\sqrt{1 - \frac{U^{2}}{v^{2}_{H}}} ~
      \cos \left (\frac{3 \alpha v_{H}}{2} ~\sqrt{\frac{a}{\mu_{\beta}}} ~
      (t - t_{M}) \right )
\nonumber \\
& &   \times ~\frac{\sin \left (\frac{3 \alpha v_{H}}{2} ~
      \sqrt{\frac{a}{\mu_{\beta}}} ~(t - t_{M}) \right )}
      {\left \vert \sin \left (\frac{3 \alpha v_{H}}{2} ~
      \sqrt{\frac{a}{\mu_{\beta}}} ~(t - t_{M}) \right ) \right \vert} ~.
\end{eqnarray}
If the eccentricity $e$ is close to $e_{1}$ $=$ 1 (see Eq. \ref{plance}),
then $I$ changes from $\sqrt{v^{2}_{H} - U^{2}}$ to
$-$ $\sqrt{v^{2}_{H} - U^{2}}$. Because of this property,
$I$ can always increase. Thus, in the planar case the orbit rotates
into position with a maximal value of $I$. As the eccentricity increases,
the dust particles gets closer to the central star, because the semi-major
axis is constant. We must note that this theory is less applicable for
larger eccentricities (see Appendix \ref{sec:applicability}).

\section{Secular evolution of Keplerian orbital elements determined
with respect to a plane perpendicular to gas flow velocity vector}
\label{sec:respect}

The choice of coordinate system was up to now arbitrary. We will denote
with two primes those coordinate systems in which the gas flow velocity
has a vector direction aligned with the direction of the $z ''$-axis.
In such a coordinate system we have $\vec{v}_{H}$ $=$ $(0,0,v_{H})$.
From Eqs. (\ref{SIC}) we obtain
\begin{eqnarray}\label{zSIC}
S &=& \sin \omega '' ~\sin i '' ~v_{H} ~,
\nonumber \\
I &=& \cos \omega '' ~\sin i '' ~v_{H} ~,
\nonumber \\
C &=& \cos i '' ~v_{H} ~.
\end{eqnarray}
From these equations we immediately obtain
\begin{equation}\label{zw}
\tan \omega '' = \frac{S}{I} ~,
\end{equation}
\begin{equation}\label{zi}
\cos i '' = \frac{C}{v_{H}} ~.
\end{equation}
The last unknown orbital element in this coordinate system is $\Omega ''$.
If we use Eqs. (\ref{zSIC}) in Eq. (\ref{dOdt}), then we can obtain
\begin{equation}\label{dzOdt1}
\frac{d \Omega ''}{dt} = - ~\frac{3 ~\alpha}{2 ~v_{H}} ~
\sqrt{\frac{p}{\mu_{\beta}}} ~\frac{S ~C}{\sin^{2} i ''} ~
\frac{e}{1 - e^{2}} ~.
\end{equation}
From the equation above we can deduce that $d \Omega ''/dt$ is always positive,
negative, or zero because $S$ and $C$ can not change sign (see Eq. \ref{S} and
Eq. \ref{C}). To find the evolution of $\Omega ''$, we divide
Eq. (\ref{dzOdt1}) by Eq. (\ref{dedt}). If we use in the result of division
Eqs. (\ref{S}), (\ref{C}), (\ref{quad1}) and (\ref{zi}), then for $I$ $\neq$
0 we finally obtain
\begin{equation}\label{dzOdt2}
\frac{d \Omega ''}{de} = \mp ~\frac{U ~V}{v^{2}_{H}} ~
\frac{e}{\sqrt{(e^{2}_{1} - e^{2})(e^{2} - e^{2}_{2})}} ~
\frac{1}{1 - e^{2} - \frac{V^{2}}{v^{2}_{H}}} ~.
\end{equation}
The minus sign is for positive values of $I$ and the plus sign is for negative
values of $I$. Integration of this equation yields
\begin{equation}\label{zOpm}
\pm ~\Omega '' = \frac{U ~V}{\vert U V \vert} ~
\arctan \sqrt{\frac{1 - \frac{V^{2}}{v^{2}_{H}} - e^{2}_{2}}
{1 - \frac{V^{2}}{v^{2}_{H}} - e^{2}_{1}} ~\frac{e^{2}_{1} - e^{2}}
{e^{2} - e^{2}_{2}}} + \psi ~.
\end{equation}
Now the plus sign is for positive values of $I$ and the minus sign is for
negative values of $I$. It is possible to show (using Eqs. \ref{S}, \ref{C}
and \ref{absI2}) that $1$ $-$ $V^{2}/v^{2}_{H}$ $-$ $e^{2}_{1}$ $\geq$ 0 and
$1$ $-$ $V^{2}/v^{2}_{H}$ $-$ $e^{2}_{2}$ $\geq$ 0. If we insert Eq. (\ref{e})
into Eq. (\ref{zOpm}), then, after some algebraic manipulations, we finally
obtain
\begin{eqnarray}\label{tanzOpm}
\tan \left (\frac{\vert U V \vert}{U V}
      \left (\pm ~\Omega '' - \psi \right ) \right ) &=& \sqrt{
      \frac{1 - \frac{V^{2}}{v^{2}_{H}} - e^{2}_{2}}
      {1 - \frac{V^{2}}{v^{2}_{H}} - e^{2}_{1}}} ~
\nonumber \\
& &   \times ~\bigg \vert \tan \bigg (\frac{3 \alpha v_{H}}{2} ~
      \sqrt{\frac{a}{\mu_{\beta}}} ~(t - t_{M}) \bigg ) \bigg \vert ~.
\end{eqnarray}
This equation leads to two equations (compare Eqs. \ref{I} and \ref{tanzOpm}).
One for $I$ $>$ 0
\begin{eqnarray}\label{tanzOp}
\tan \left (\Omega '' - \psi_{+} \right ) &=& - ~
      \frac{U ~V}{\vert U V \vert} ~
      \sqrt{\frac{1 - \frac{V^{2}}{v^{2}_{H}} - e^{2}_{2}}
      {1 - \frac{V^{2}}{v^{2}_{H}} - e^{2}_{1}}} ~
\nonumber \\
& &   \times ~\tan \left (\frac{3 \alpha v_{H}}{2} ~
      \sqrt{\frac{a}{\mu_{\beta}}} ~(t - t_{M}) \right )
\end{eqnarray}
and one for $I$ $<$ 0
\begin{eqnarray}\label{tanzOm}
\tan \left (- \Omega '' - \psi_{-} \right ) &=& ~
      \frac{U ~V}{\vert U V \vert} ~
      \sqrt{\frac{1 - \frac{V^{2}}{v^{2}_{H}} - e^{2}_{2}}
      {1 - \frac{V^{2}}{v^{2}_{H}} - e^{2}_{1}}} ~
\nonumber \\
& &   \times ~\tan \left (\frac{3 \alpha v_{H}}{2} ~
      \sqrt{\frac{a}{\mu_{\beta}}} ~(t - t_{M}) \right ) ~.
\end{eqnarray}
These solutions meet at time $t_{M}$ when $I$ changes its sign
(see Eq. \ref{I}). Close to the time $t_{M}$, the right-hand side of both
equations is close to zero. For values of $\psi_{+}$ and $\psi_{-}$ at
the time $t_{M}$ we obtain
\begin{equation}\label{zOpsip}
\psi_{+} = \Omega_{M} '' - k_{3} ~\pi ~,
\end{equation}
\begin{equation}\label{zOpsim}
\psi_{-} = - ~\Omega_{M} '' - k_{4} ~\pi ~,
\end{equation}
where $k_{3}$ and $k_{4}$ are two integers and $\Omega_{M} ''$ is
the value of $\Omega ''$ close to the time $t_{M}$. In Eqs. (\ref{tanzOp})
and (\ref{tanzOm}) both these values lead to the same solution
\begin{eqnarray}\label{zO}
\tan \left (\Omega '' - \Omega_{M} '' \right ) &=& - ~
      \frac{U ~V}{\vert U V \vert} ~
      \sqrt{\frac{1 - \frac{V^{2}}{v^{2}_{H}} - e^{2}_{2}}
      {1 - \frac{V^{2}}{v^{2}_{H}} - e^{2}_{1}}} ~
\nonumber \\
& &   \times ~\tan \left (\frac{3 \alpha v_{H}}{2} ~
      \sqrt{\frac{a}{\mu_{\beta}}} ~(t - t_{M}) \right ) ~.
\end{eqnarray}
This equation is a generalization of the result of \cite{bera}.
From the discussion of Eq. (\ref{dzOdt1}) we know that $\Omega ''$ is
a monotonic or constant (for $S$ $=$ 0 or $C$ $=$ 0 and $e$ $\neq$ 0 or
$e$ $\neq$ 1) function of time. If we consider values of $\Omega ''$ only
in the interval [0, $2 \pi$), then $\Omega ''$ is a periodic or constant
function of time.

\subsection{Stationary solution}
\label{subsec:stationary}

Eqs. (\ref{dSdt})--(\ref{dCdt}) and (\ref{S})--(\ref{absI2}) enable finding
a stationary solution determined by two equations
\begin{equation}\label{statsol1}
I = 0
\end{equation}
and
\begin{equation}\label{statsol2}
\frac{eC^{2}}{1 - e^{2}} - \frac{S^{2}}{e} = 0 ~.
\end{equation}
If we insert Eqs. (\ref{S}) and (\ref{C}) into Eq. (\ref{statsol2}),
we obtain a condition for the eccentricity of the stationary solution
\begin{equation}\label{statsole}
e^{2} = \frac{\vert U \vert}{\vert U \vert + \vert V \vert} ~.
\end{equation}
If Eqs. (\ref{statsol1}) and (\ref{statsole}) are fulfilled, then
$e$, $S$, $I$ and $C$ remain constant.

For orbital elements determined with respect to the plane perpendicular to
the gas flow velocity vector, we obtain from condition $I$ $=$ 0 in the second
of Eqs. (\ref{zSIC}) that $\cos \omega ''$ $=$ 0 (since $\sin i''$ $\neq$ 0).
Hence
\begin{equation}\label{statsolzw}
\omega '' = \frac{\pi}{2} + k_{5} \pi ~,
\end{equation}
where $k_{5}$ is an integer. Eq. (\ref{didt}) yields that the inclination
is in this case a constant. Finally from Eq. (\ref{dOdt}) we obtain that
$\Omega ''$ depend linearly on time. $\Omega ''$ is the only orbital element
of the stationary solution that varies with time. Therefore, the orbital
plane rotates around a line aligned with the gas flow velocity vector and
going through the centre of gravity. This stationary solution is in
accordance with the result of \cite{bera}.

\section{Evolution of orbital elements determined with respect to an arbitrary
reference plane}
\label{sec:arbitrary}

It is useful to have the evolution of orbital elements in a reference frame
oriented arbitrarily with respect to the hydrogen gas velocity vector.
Therefore in this section we transform the orbital elements derived in
the previous section into such a frame.

\begin{figure}[t]
\begin{center}
\includegraphics[height=0.23\textheight]{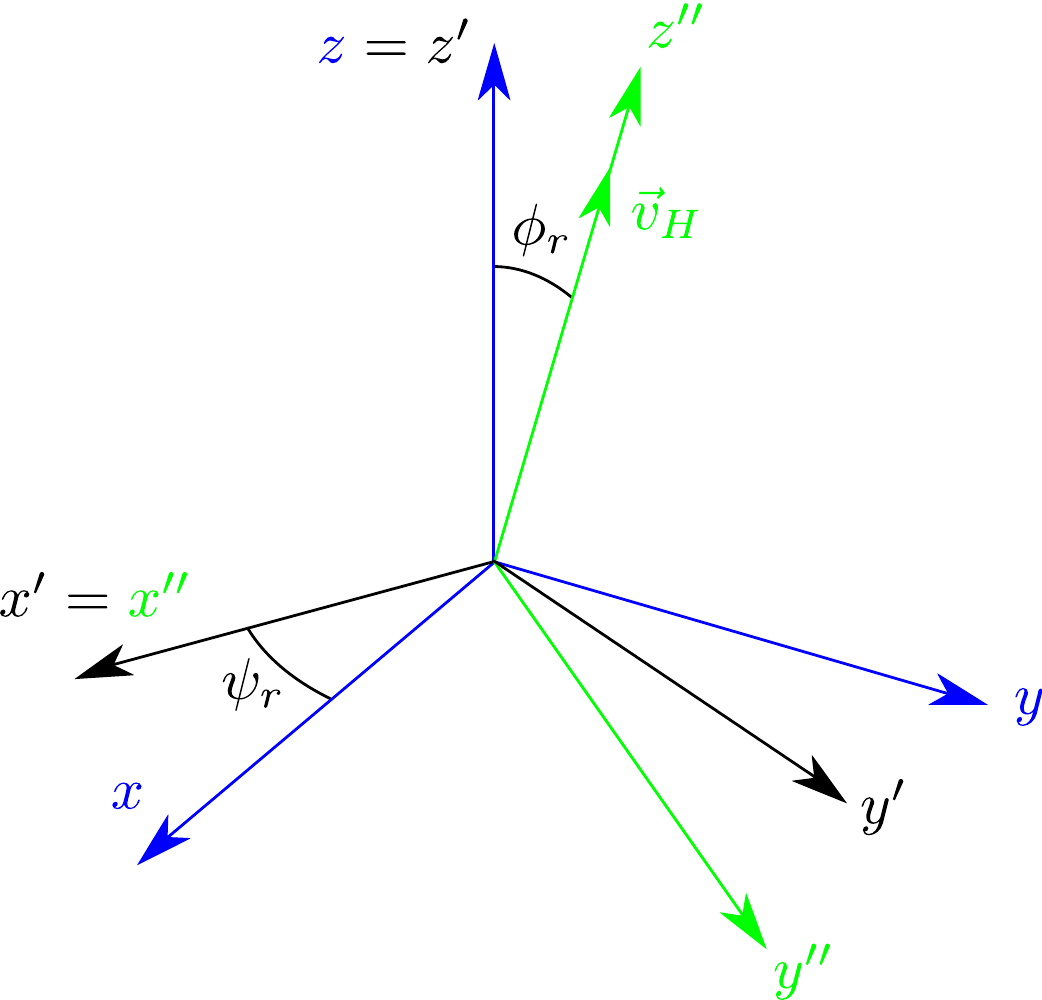}
\end{center}
\caption{Transformation from a frame in which the $z''$-axis is aligned
with the interstellar gas velocity vector into an arbitrary
reference frame.}
\label{F2}
\end{figure}

The frame denoted with two primes has the $z ''$-axis aligned with
the velocity vector of the hydrogen gas. $x''$ is the axis from which
$\Omega ''$ is measured. For the sake of simplicity we will
assume that the $x ''$-axis lies in the $xy$-plane of the frame into
which we want to transform the orbital elements. By making this
assumption we do not lose any generality of the solved problem since
the direction of the $x''$-axis is arbitrary. The transformation of the
particle's coordinates form the two primed frame into the unprimed frame
can be made by the composition of two rotations (see Fig. \ref{F2}).
The first rotation is around the $x ''$-axis by the angle between
$z '$ and $z ''$.
\begin{eqnarray}\label{rot1}
x ' &=& x '' ~,
\nonumber \\
y ' &=& y '' \cos \phi_{r} + z '' \sin \phi_{r} ~,
\nonumber \\
z ' &=& - ~y '' \sin \phi_{r} + z '' \cos \phi_{r} ~,
\end{eqnarray}
where
\begin{eqnarray}\label{rot1angle}
\sin \phi_{r} &=& \frac{\sqrt{v^{2}_{HX} + v^{2}_{HY}}}{v_{H}} ~,
\nonumber \\
\cos \phi_{r} &=& \frac{v_{HZ}}{v_{H}} ~.
\end{eqnarray}
The second rotation is a rotation
around the $z '$-axis by the angle between $x$ and $x '$.
\begin{eqnarray}\label{rot2}
x &=& x ' \cos \psi_{r} + y ' \sin \psi_{r} ~,
\nonumber \\
y &=& - ~x ' \sin \psi_{r} + y ' \cos \psi_{r} ~,
\nonumber \\
z &=& z ' ~,
\end{eqnarray}
where
\begin{eqnarray}\label{rot2angle}
\sin \psi_{r} &=& \frac{v_{HX}}{\sqrt{v^{2}_{HX} + v^{2}_{HY}}} ~,
\nonumber \\
\cos \psi_{r} &=& \frac{v_{HY}}{\sqrt{v^{2}_{HX} + v^{2}_{HY}}} ~.
\end{eqnarray}
Hence the transformation from the two primed frame into the unprimed
frame is
\begin{eqnarray}\label{transformation}
x &=& x '' \cos \psi_{r} + (y '' \cos \phi_{r} + z '' \sin \phi_{r})
      \sin \psi_{r} ~,
\nonumber \\
y &=& - ~x '' \sin \psi_{r} + (y '' \cos \phi_{r} + z '' \sin \phi_{r})
      \cos \psi_{r} ~,
\nonumber \\
z &=& - ~y '' \sin \phi_{r} + z '' \cos \phi_{r} ~,
\end{eqnarray}
In order to find $\Omega$ and $i$ we can transform the coordinates
of the unit vector $\vec{e} ''_{N}$ $=$ ($\sin \Omega ''$ $\sin i ''$,
$- \cos \Omega ''$ $\sin i ''$, $\cos i ''$).
We obtain
\begin{eqnarray}\label{transzen}
e_{NX} &=& \sin \Omega '' \sin i '' ~
      \frac{v_{HY}}{\sqrt{v^{2}_{HX} + v^{2}_{HY}}}
\nonumber \\
& &   + ~\left (- \cos \Omega '' \sin i '' ~\frac{v_{HZ}}{v_{H}} + \cos i '' ~
      \frac{\sqrt{v^{2}_{HX} + v^{2}_{HY}}}{v_{H}} \right )
\nonumber \\
& &   \times ~\frac{v_{HX}}{\sqrt{v^{2}_{HX} + v^{2}_{HY}}} ~,
\nonumber \\
e_{NY} &=& - \sin \Omega '' \sin i '' ~
      \frac{v_{HX}}{\sqrt{v^{2}_{HX} + v^{2}_{HY}}}
\nonumber \\
& &   + ~\left (- \cos \Omega '' \sin i '' ~\frac{v_{HZ}}{v_{H}} + \cos i '' ~
      \frac{\sqrt{v^{2}_{HX} + v^{2}_{HY}}}{v_{H}} \right )
\nonumber \\
& &   \times ~\frac{v_{HY}}{\sqrt{v^{2}_{HX} + v^{2}_{HY}}} ~,
\nonumber \\
e_{NZ} &=& \cos \Omega '' \sin i '' ~
      \frac{\sqrt{v^{2}_{HX} + v^{2}_{HY}}}{v_{H}} +
      \cos i '' ~\frac{v_{HZ}}{v_{H}} ~.
\end{eqnarray}
In this equation, Eq. (\ref{zO}) and Eq. (\ref{zi}) have to be used
to determine the values of $\Omega ''$ and $i ''$. Since $\vec{e}_{N}$ $=$
($\sin \Omega$ $\sin i$, $- \cos \Omega$ $\sin i$, $\cos i$), we can calculate
$\Omega$ and $i$ from
\begin{equation}\label{O}
\tan \Omega = - ~\frac {e_{NX}}{e_{NY}} ~,
\end{equation}
\begin{equation}\label{i}
\cos i = e_{NZ} ~.
\end{equation}

Now we can find $\omega$ from Eqs. (\ref{SIC}). One can easily verify that
\begin{eqnarray}\label{J}
J = S \cos \omega - I \sin \omega &=& \cos \Omega ~v_{HX} +
      \sin \Omega ~v_{HY} ~,
\end{eqnarray}
\begin{eqnarray}\label{H}
H = S \sin \omega + I \cos \omega &=& - ~\sin \Omega ~\cos i ~v_{HX}
\nonumber \\
& &   + ~\cos \Omega ~\cos i ~v_{HY}
\nonumber \\
& &   + ~\sin i ~v_{HZ} ~.
\end{eqnarray}
Therefore
\begin{equation}\label{w}
\sin \omega = \frac{HS - JI}{S^{2} + I^{2}} ~~\mbox{and}~~
\cos \omega = \frac{JS + HI}{S^{2} + I^{2}} ~.
\end{equation}

\subsection{Perpendicular case}
\label{sec:perpendicular}

If the velocity vector of a neutral gas is perpendicular to the line
of apsides, then $S$ $=$ 0 ($U$ $=$ 0) and the value of $S$ does
not change with time. For this special case we obtain from
Eqs. (\ref{e1})--(\ref{e2})
\begin{equation}\label{perce1}
e^{2}_{1} = 1 - \frac{V^{2}}{v^{2}_{H}} ~,
\end{equation}
\begin{equation}\label{perce2}
e^{2}_{2} = 0 ~.
\end{equation}
Therefore the minimum eccentricity is 0 and the maximum eccentricity is
$\sqrt{1 - V^{2}/v^{2}_{H}}$. Using Eqs. (\ref{perce1})--(\ref{perce2}) we
obtain from Eqs. (\ref{C}), (\ref{e}) and (\ref{I}) for the perpendicular
case
\begin{equation}\label{percC}
C = \frac{V}{\sqrt{1 - e^{2}}} ~,
\end{equation}
\begin{equation}\label{percI}
I = - ~\frac{v_{H}}{e ~\sqrt{1 - e^{2}}} ~
\frac{1 - \frac{V^{2}}{v^{2}_{H}}}{2} ~
\sin \left (3 \alpha v_{H} ~\sqrt{\frac{a}{\mu_{\beta}}} ~
(t - t_{M}) \right ) ~,
\end{equation}
\begin{equation}\label{perce}
e^{2} = \left (1 - \frac{V^{2}}{v^{2}_{H}} \right )
\cos^{2} \left (\frac{3 \alpha v_{H}}{2} ~\sqrt{\frac{a}{\mu_{\beta}}} ~
(t - t_{M}) \right ) ~.
\end{equation}
We mention that $C^{2}$ $+$ $I^{2}$ $=$ $v^{2}_{H}$ always holds
for the perpendicular case.

For the inclination in the two primed frame, we have from Eq. (\ref{C}) and
Eq. (\ref{zi})
\begin{equation}\label{perczi}
\cos i '' = \frac{V}{v_{H} \sqrt{1 - e^{2}}} ~.
\end{equation}
Hence for $i ''$ we obtain that $i ''$ $\in$ (0, $\arccos (V/v_{H})$)
for positive values of $V$ and $i ''$ $\in$ ($\arccos (V/v_{H})$, $\pi$)
for negative values of $V$.

Since $S$ $=$ 0 and $\sin i ''$ $\neq$ 0, we obtain from
the first of Eqs. (\ref{zSIC}) that $\sin \omega ''$ $=$ 0. Therefore
\begin{equation}\label{perczw}
\omega '' = k_{6} \pi ~,
\end{equation}
where $k_{6}$ is an integer.

Finally Eq. (\ref{dzOdt1}) for $\sin i ''$ $\neq$ 0 yields that $\Omega ''$
is a constant.

Now we find relations between the orbital elements in the unprimed frame
for the special case $S$ $=$ 0. If we combine Eq. (\ref{dwdt}) for $S$ $=$ 0
with Eq. (\ref{didt}), we get
\begin{equation}\label{tilteq}
\frac{\cos \omega}{\sin \omega} ~d \omega = - ~\frac{\cos i}{\sin i} ~di
\end{equation}
which has the solution
\begin{equation}\label{abstilt}
\vert \sin i \vert ~\vert \sin \omega \vert = d ~,
\end{equation}
where $d$ is an integration constant. Since $i$ $\in$ (0, $\pi$),
$\sin i$ $>$ 0. If $\sin \omega$ $\neq$ 0 initially, then $d$ $\neq$ 0
and therefore $\sin \omega$ can not change sign. Hence
\begin{equation}\label{tilt}
\sin i ~\sin \omega = l ~,
\end{equation}
where $l$ is a constant. From Eq. (\ref{tilt}) we can see
that the orbital plane is only tilted back and forth around
the line of apsides for the special case $S$ $=$ 0.
The instantaneous value of $i$ can be determined from Eq. (\ref{i}).
In order to find $\Omega$ we combine Eq. (\ref{dOdt}) with Eq. (\ref{didt}).
We get differential equation
\begin{equation}\label{tiltOeq}
\cos \omega ~d \Omega= \frac{\sin \omega}{\sin i} ~di ~.
\end{equation}
Equation (\ref{tilt}) allows us to convert this expression into an equation
for $\Omega (i)$ with solution
\begin{equation}\label{tiltO}
\tan (\Omega - \Omega_{0}) = - ~\frac{\sin \omega ~\cos i}{\cos \omega} ~,
\end{equation}
where $\Omega_{0}$ is the value of $\Omega$ for $i$ $=$ $\pi/2$.

\section{Conclusion}
\label{sec:conclusion}

We have investigated the long-term orbital evolution of a spherical
dust particle perturbed by a small constant mono-directional
force caused by a fast interstellar gas flow. The secular time derivatives of
the particle's Keplerian orbital elements were derived. We
transformed the system of differential equations for the Keplerian
orbital elements into a system of differential equations for the
semi-major axis, eccentricity, and the radial, transversal, and normal
components of the interstellar gas flow velocity vector determined at the
pericentre of the particle's orbit. In these new variables, the system of
differential equations can be easily solved. The solution of the system
was used in order to obtain the evolution of the Keplerian
orbital elements in a reference frame in which the orbital
elements are determined with respect to the plane perpendicular
to the interstellar gas velocity vector. We found an explicit form for
the time dependence of all Keplerian orbital elements in this reference
frame. We generalized the expression for the time dependence of the longitude
of the ascending node found by \cite{bera}. We transformed newly found
orbital elements into an arbitrary reference frame. This transformation gave
us explicit time dependences of all Keplerian orbital elements in
the arbitrary reference frame. The orbital elements of the dust
particle in an arbitrary reference frame change periodically with
a constant oscillation period or else remain constant. We determined
the properties of the solution for the planar case and for the case in
which the gas flow velocity vector is perpendicular to the line
of apsides. We found the maximal and minimal values of the eccentricity
in these cases. In the planar case, the particle's orbit approaches
the direction with maximal value of $I$. In the perpendicular case,
the orbital plane is tilted back and forth around the line of apsides.
We also confirmed the stationary solution found by \cite{bera}.
For the stationary solution, the orbital plane rotates around the line aligned
with the gas flow velocity vector and going through the centre of gravity.

This solution can be applied also for the dust particles in the Solar
system. If we consider icy particles with radii from 1 to 10 $\mu$m,
then the solution is valid for orbits with semi-major axis from 200 to
3000 AU, approximately. More exact values of these limits depend on the radius
of the particle. A maximal orbit eccentricity for which the solution is valid
is smaller than approximately 0.8 and depends on the semi-major axis
of the orbit and the particle's radius (see Fig. \ref{F3}).
The period of change of the orbital elements for these orbits ranges from
10$^{5}$ to 2 $\times$ 10$^{6}$ years, approximately.

\appendix

\section{Applicability of assumptions (f)--(h) in Solar system}
\label{sec:applicability}

\begin{figure*}
\begin{center}
\includegraphics[width=\textwidth]{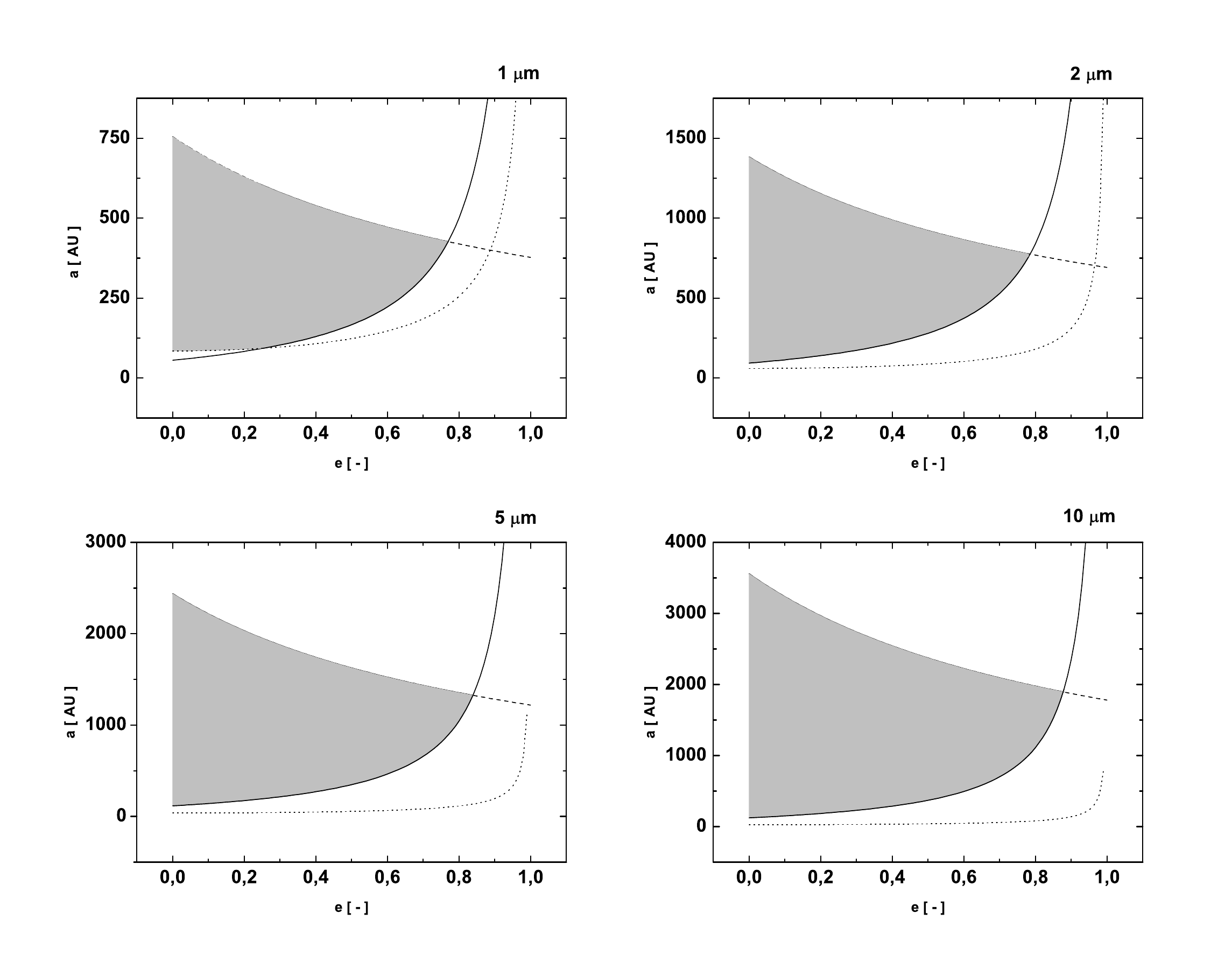}
\end{center}
\caption{Allowed values of orbital semi-major axes and eccentricities
in the $ae$-phase plane determined using the relations between the semi-major
axis and eccentricity obtained from (\ref{A2}), (\ref{A4}) and
(\ref{A5}). Dust particles have radius $R$ $\in$ \{1 $\mu$m, 2 $\mu$m,
5 $\mu$m, 10 $\mu$m\} and mass density $\varrho$ $=$ 1 g cm$^{-3}$ and
$\bar{Q}'_{pr} = 1$. Dashed line indicates
(\ref{A2}), solid line, (\ref{A4}) and dotted line,
(\ref{A5}). The allowed region of the semi-major axes and
eccentricities is depicted in grey.}
\label{F3}
\end{figure*}

Assumption (h) can be mathematically written as
\begin{equation}\label{A1}
\frac{\mu_{\beta}}{r^{2}} \gg \frac{c_{D} ~n_{H} ~m_{H} ~A ~v_{H}^{2}}{m} ~.
\end{equation}
This inequality must always hold during orbital evolution. The right-hand side
of the inequality is a constant and the left-hand side is minimal
at maximal $r$. Since we consider elliptical orbits, $r$ is maximal at
the apocentre of the orbit. Therefore we will check this inequality at
the apocentre. In the apocentre we can rewrite (\ref{A1}) as
\begin{equation}\label{A2}
\frac{\mu_{\beta}}{a^{2} ~(1 + e)^{2}} \gg
\frac{3 ~c_{D} ~n_{H} ~m_{H} ~v_{H}^{2}}{4 ~R ~\varrho} ~,
\end{equation}
where $a$ is the semi-major axis and $e$ is the eccentricity of
the dust particle's elliptical orbit. For a given particle this inequality
determines the allowed $a$ and $e$ in the $ae$-phase plane.

Assumption (f) can be mathematically written as
\begin{equation}\label{A3}
\vert \vec{v}_{H} \vert \gg \vert \vec{v} \vert ~.
\end{equation}
Because the velocity of the particle is maximal at the pericentre of the orbit,
we will check this inequality at the pericentre. In the pericentre
we can rewrite (\ref{A3}) as
\begin{equation}\label{A4}
\vert \vec{v}_{H} \vert \gg
\sqrt{\frac{\mu_{\beta}}{a} ~\frac{1+ e}{1 - e}} ~.
\end{equation}
Similarly to (\ref{A2}), this inequality also determines
the allowed $a$ and $e$ in the $ae$-phase plane.

Assumption (g) can be approximately mathematically written as
(\citealt{wywh})
\begin{equation}\label{A5}
- \left \langle \frac{da}{dt} \right \rangle_{PR} ~\frac{1}{a} =
\frac{\beta ~\mu}{c ~a^{2}} ~\frac{2 + 3 e^{2}}{(1 - e^{2})^{3/2}} \ll
b_{a} ~,
\end{equation}
and
\begin{equation}\label{A6}
- \left \langle \frac{de}{dt} \right \rangle_{PR} ~\frac{1}{e} =
\frac{5 ~\beta ~\mu}{2 ~c ~a^{2}} ~\frac{1}{(1 - e^{2})^{1/2}} \ll
b_{e} ~.
\end{equation}
Here, $b_{a}$ and $b_{e}$ determine the maximal values for the relative secular
time derivatives. Inequalities (\ref{A5}) and (\ref{A6}) for a given particle,
$b_{a}$ and $b_{e}$ determine the allowed $a$ and $e$ in the $ae$-phase plane.
It is possible to show that for $b_{e}$ $>$ $5 b_{a} / 4$,
the semi-major axis determined from (\ref{A5}) is always greater than
the semi-major axis determined from (\ref{A6}) for $e$ $\in$ $[0,1)$.
Therefore we will check only inequality (\ref{A5}). Using
inequality (\ref{A5}) we take into account also the eccentricity of
the particle's orbits. This is a more general method than the method used
by \cite{bera} who considered only circular orbits (see Eq. 26
in \citealt{bera}).

To obtain the conditions in the Solar system we used
$L_{\odot}$ $=$ 3.842 $\times$ 10$^{26}$ W (\citealt{bahcall}),
$c_D$ $=$ 2.6 (\citealt{bana,scherer,flow}),
$n_{H}$ $=$ 0.2 cm$^{-3}$ (\citealt{bera}) and
$v_{H}$ $=$ 26 km s$^{-1}$ (e.g. \citealt{lallement,landgraf}).
In Fig. \ref{F3} four panels are depicted. Each of these corresponds
to a different particle radius. We used particles with radius $R$ $\in$
\{1 $\mu$m, 2 $\mu$m, 5 $\mu$m, 10 $\mu$m\} and mass density $\varrho$ $=$
1 g cm$^{-3}$ and $\bar{Q}'_{pr}$ $=$ 1. Dependences between the semi-major
axis and eccentricity are obtained from (\ref{A2}), (\ref{A4})
and (\ref{A5}). In (\ref{A2}) we assumed that the left-hand side must
be at least 10 times greater than the right-hand side. For the equation
obtained using this inequality we used the dashed line in Fig. \ref{F3}.
In (\ref{A4}) we also assumed that left-hand side must be at
least 10 times greater than the right-hand side. In Fig. \ref{F3}
this is depicted using a solid line. In (\ref{A5}) we assumed
that $b_{a}$ $=$ 10$^{-6}$ year$^{-1}$ and the right-hand side must
also be at least 10 times greater than the left-hand side. In Fig. \ref{F3}
(\ref{A5}) is depicted useing a dotted line. The grey region in Fig. \ref{F3}
represents values of semi-major axes and eccentricities in the $ae$-phase plane
for which all three inequalities hold. In Fig. \ref{F3} we can see that
assumptions (f)--(h) can be applied to the Solar system. If also
assumptions (a)--(e) hold, then Eq. (\ref{fineqom}) can be used.
Assumption (h) enables the use of perturbation theory.

\section{Connection with previous work}
\label{sec:connection}

\cite{flow} take into account also the relative velocity
$\vec{v}$ of the dust particle with respect to the star in the force
caused by the neutral interstellar gas. The particle's equation of motion has
the following form
\begin{equation}\label{B1}
\frac{d \vec{v}}{dt} = - ~\frac{\mu_{\beta}}{r^{3}} \vec{r} -
c_{D} ~\gamma_{H} ~ \vert \vec{v} - \vec{v}_{H} \vert ~
\left ( \vec{v} - \vec{v}_{H} \right ) ~.
\end{equation}
After expansion of the right-hand side of this equation into a series
the linear term in the relative velocity is included in the calculation
of the secular time derivatives of the Keplerian orbital elements.
The result for the secular time derivative of the semi-major axis is
\begin{eqnarray}\label{B2}
\left \langle \frac{da}{dt} \right \rangle &=& - ~2 ~a ~c_{D} ~\gamma_{H} ~
      v_{H}^{2} ~\sqrt{\frac{p}{\mu_{\beta}}} ~\sigma ~
      \Biggl \{1 + \frac{1}{v_{H}^{2}}
\nonumber \\
& &   \times ~\Biggl [I^{2} - (I^{2} - S^{2})
      \frac{1 - \sqrt{1 - e^{2}}}{e^{2}} \Biggr ] \Biggr \} ~,
\end{eqnarray}
where
\begin{equation}\label{B3}
\sigma = \frac{\sqrt{\mu_{\beta}/p}}{v_{H}} ~.
\end{equation}
From Eq. (\ref{B2}) we can see that the secular time derivative of
the semi-major axis is proportional to the value of the semi-major axis
(the value of $\sqrt{p/ \mu_{\beta}} ~\sigma$ is independent of
the semi-major axis). The term within the square brackets ($SB$)
in Eq. (\ref{B2}) can be written as
\begin{equation}\label{B4}
SB = I^{2} - (I^{2} - S^{2}) \frac{1 - \sqrt{1 - e^{2}}}{e^{2}} =
\frac{1 - \sqrt{1 - e^{2}}}{e^{2}} ~(I^{2} \sqrt{1 - e^{2}} + S^{2}) \geq 0 ~.
\end{equation}
The secular semi-major axis is a decreasing function of time.

Now, we find the maximal possible decrease of the secular semi-major axis.
Because terms the multiplied by $S^{2}$ and $I^{2}$ are both positive, we
obtain the maximal value of $SB$ for the orbit orientation characterized by
$C$ $=$ 0. If $C$ $=$ 0, then $S^{2}$ $+$ $I^{2}$ $=$ $v_{H}^{2}$.
Using this, the value of $SB$ can be written as
\begin{equation}\label{B5}
SB = \frac{1 - \sqrt{1 - e^{2}}}{e^{2}} ~
[v_{H}^{2} \sqrt{1 - e^{2}} + S^{2} (1 - \sqrt{1 - e^{2}})] ~.
\end{equation}
Here, $v_{H}^{2}$ is constant. Therefore, we obtain the maximal value of $SB$
for the orbit orientation characterized by $S^{2}$ $=$ $v_{H}^{2}$.
Hence, the maximal value of $SB$ is
\begin{equation}\label{B6}
SB = v_{H}^{2} ~\frac{1 - \sqrt{1 - e^{2}}}{e^{2}} = v_{H}^{2} ~g(e) ~.
\end{equation}
In order to find the behaviour of the function $g(e)$ we can write
\begin{equation}\label{B7}
\frac{dg(e)}{de} = \frac{2 - e^{2} - 2 \sqrt{1 - e^{2}}}
{e^{3} \sqrt{1 - e^{2}}} ~,
\end{equation}
\begin{eqnarray}\label{B8}
\frac{dg_{1}(e)}{de} &=& \frac{d}{de} ~
      \bigl (2 - e^{2} - 2 \sqrt{1 - e^{2}} \bigr )
\nonumber \\
&=&   - 2 ~e + \frac{2 ~e}{\sqrt{1 - e^{2}}} \geq 0 ~.
\end{eqnarray}
Because $dg_{1}(e)/de$ $\geq$ 0, $g_{1}(e)$ is an increasing function
of the eccentricity. The value of $g_{1}(0)$ is 0. Therefore, $g_{1}(e)$ is
positive for $e$ $\in$ (0, 1]. If $g_{1}(e)$ is positive, then
$dg(e)/de$ $>$ 0. Because $dg(e)/de$ $>$ 0, the function $g(e)$
is an increasing function of the eccentricity for $e$ $\in$ (0, 1].
The function $g(e)$ has its maximal value for $e$ $=$ 1. Therefore the maximal
value of $SB$ is
\begin{equation}\label{B9}
SB_{\mbox{max}} = v_{H}^{2} ~.
\end{equation}
Hence, the maximal possible decrease of the secular semi-major axis is
\begin{equation}\label{B10}
\left \langle \frac{da}{dt} \right \rangle_{\mbox{max}} = - ~4 ~
a ~c_{D} ~\gamma_{H} ~
v_{H}^{2} ~\sqrt{\frac{p}{\mu_{\beta}}} ~\sigma ~.
\end{equation}

\begin{table}[t]
\centering
\begin{tabular}{| c | c |}
\hline
R & $T_{a}$\\

[$\mu$m] & $\left [10^{3} \atop \mbox{years} \right ]$\\
\hline
1 & 46.7\\
2 & 93.4\\
5 & 233.4\\
10 & 466.8\\
\hline
\end{tabular}
\caption{Time intervals $T_{a}$ after which the decrease of the semi-major
axis is smaller than 10\% ($c_{a}$ $=$ 0.9) for particles with
$R$ $\in$ \{1 $\mu$m, 2 $\mu$m, 5 $\mu$m, 10 $\mu$m\} and mass density
$\varrho$ $=$ 1 g cm$^{-3}$ in the Solar system.}
\label{T2}
\end{table}

We define a time interval $T_{a}$ during which we suppose that the semi-major
axis is approximately constant. For a given value of the semi-major axis
we have
\begin{equation}\label{B11}
\frac{a + \left \langle \frac{da}{dt} \right \rangle_{\mbox{max}} ~
T_{a}}{a} = \mbox{constant} = c_{a} \approx 1 ~.
\end{equation}
Because we use the maximal possible decrease of the secular semi-major axis,
the left-hand side of Eq. (\ref{B11}) will be even closer to unity for the real
secular time derivative of the semi-major axis.
However
\begin{equation}\label{B12}
\frac{T_{a}}{T_{e}} = \frac{3 ~T_{a} ~\alpha ~v_{H}}{2 ~\pi} ~
\sqrt{\frac{a}{\mu_{\beta}}}
\end{equation}
(see Eq. \ref{Te}) is proportional to $a^{1/2}$. Therefore, for larger
semi-major axes we obtain more eccentricity periods in the same time
interval $T_{a}$. Therefore the theory with the equation of motion
considered in the form of Eq. (\ref{fineqom}) is better applicable
for larger semi-major axes. We can calculate the values of $T_{a}$ from
Eq. (\ref{B11}). If we use $c_{a}$ $=$ 0.9, then we obtain for dust particles
with $R$ $\in$ \{1 $\mu$m, 2 $\mu$m, 5 $\mu$m, 10 $\mu$m\} and mass density
$\varrho$ $=$ 1 g cm$^{-3}$ in the Solar system (see Appendix
\ref{sec:applicability}) the values of $T_{a}$ shown in Table \ref{T2}.

\begin{acknowledgements}
This paper was supported by Scientific Grant Agency VEGA
grant No. 2/0016/09.
\end{acknowledgements}


\begin{thebibliography}{99}

\bibitem[Alouani-Bibi et al.(2011)]{alouani}Alouani-Bibi, F., Opher, M.,
Alexashov, D., Izmodenov, V., Toth, G.: Kinetic versus multi-fluid
approach for interstellar neutrals in the heliosphere: Exploration of
the interstellar magnetic field effects. Astrophys. J. {\bf 734}, 45 (2011)

\bibitem[Bahcall(2002)]{bahcall}Bahcall, J.: The luminosity
constraint on solar neutrino fluxes. Phys. Rev. C {\bf 65}, 025801
(2002)

\bibitem[Baines et al.(1965)]{baines}Baines, M.J., Williams, I.P.,
Asebiomo, A.S.: Resistance to the motion of a small sphere
moving through a gas. Mon. Not. R. Astron. Soc. {\bf 130}, 63--74 (1965)

\bibitem[Banaszkiewicz et al.(1994)]{bana}Banaszkiewicz, M.,
Fahr, H.J., Scherer, K.: Evolution of dust particle orbits under
the influence of solar wind outflow asymmetries and the formation
of the zodiacal dust cloud. Icarus {\bf 107}, 358--374 (1994)

\bibitem[Belyaev and Rafikov(2010)]{bera}Belyaev, M., Rafikov, R.:
The dynamics of dust grains in the outer Solar System.
Astrophys. J. {\bf 723}, 1718--1735 (2010)

\bibitem[Burns et al.(1979)]{burns}Burns, J.A., Lamy, P.L., Soter, S.:
Radiation forces on small particles in the Solar System.
Icarus {\bf 40}, 1--48 (1979)

\bibitem[Danby(1988)]{danby}Danby, J.M.A.: Fundamentals
of Celestial Mechanics, 2nd edn. Willmann-Bell, Richmond, VA, USA (1988)

\bibitem[Debes et al.(2009)]{debes}Debes, J.H., Weinberger, A.J.,
Kuchner, M.J.: Interstellar medium sculpting of the HD 32297
debris disk. Astrophys. J. {\bf 702}, 318--326 (2009)

\bibitem[Dermott et al.(1994)]{dermott}Dermott, S.F., Jayaraman, S.,
Xu, Y.L., Gustafson, B.\AA.S., Liou, J.C.: A circumsolar ring
of asteroidal dust in resonant lock with the Earth. Nature
{\bf 369}, 719--723 (1994)

\bibitem[Fahr(1996)]{fahr}Fahr, H.J.: The interstellar gas flow
through the heliospheric interface region. Space Sci. Rev. {\bf 78},
199--212 (1996)

\bibitem[Hines et al.(2007)]{hines}Hines, D.C., Schneider, G., Hollenbach, D.,
Mamajek, E.E., Hillenbrand, L.A., Metchev, S.A., Meyer, M.R., Carpenter, J.M.,
Moro-Mart\'{i}n, A., Silverstone, M.D., Kim, J.S., Henning, T., Bouwman, J.,
Wolf, S.: The Moth: An unusual circumstellar structure associated
with HD 61005. Astrophys. J. {\bf 671}, L165--L168 (2007)

\bibitem[Kimura and Mann(1998)]{kimura}Kimura, H., Mann, I.:
The electric charging of interstellar dust in the solar system and
consequences for its dynamics. Astrophys. J. {\bf 499}, 454--462 (1998)

\bibitem[Kla\v{c}ka(2008)]{klacka}Kla\v{c}ka, J.: Electromagnetic
radiation, motion of a particle and energy--mass relation.
arXiv: astro-ph/0807.2915 (2008)

\bibitem[Kla\v{c}ka et al.(2009a)]{NOVA}Kla\v{c}ka, J., K\'{o}mar, L.,
P\'{a}stor, P., Petr\v{z}ala, J.: Solar wind and motion of
interplanetary dust grains. In: Johannson, H.E. (ed.) Handbook
on Solar Wind: Effects, Dynamics and Interactions, pp. 227--273. NOVA Science
Publishers, New York (2009a)

\bibitem[Kla\v{c}ka et al.(2009b)]{SW}Kla\v{c}ka, J., Petr\v{z}ala, J.,
P\'{a}stor, P., K\'{o}mar, L.: Solar wind and motion of dust grains.
arXiv: astro-ph/0904.2673 (2009b)

\bibitem[Lallement(1996)]{lallement}Lallement, R.: Relations
between ISM inside and outside the heliosphere. Space Sci. Rev.
{\bf 78}, 361--374 (1996)

\bibitem[Landgraf et al.(1999)]{landgraf}Landgraf, M., Augustsson, K.,
Gr\"{u}n, E., Gustafson, B.\AA.S.: Deflection of the local
interstellar dust flow by solar radiation pressure. Science
{\bf 286}, 2319--2322 (1999)

\bibitem[Lantoine and Russell(2011)]{lantoine}Lantoine, G., Russell, R.P.:
Complete closed-form solutions of the Stark problem.
Celest. Mech. Dyn. Astron. {\bf 109}, 333--366 (2011)

\bibitem[Marzari and Th\'{e}bault(2011)]{marzari}Marzari, F., Th\'{e}bault, P.:
On how optical depth tunes the effects of the interstellar medium on
debris discs. Mon. Not. R. Astron. Soc. (2011).
doi: 10.1111/j.1365-2966.2011.19161.x

\bibitem[Mignard and Henon(1984)]{mihe}Mignard, F., Henon, M.:
About an unsuspected integrable problem. Celest. Mech. Dyn. Astron.
{\bf 33}, 239--250 (1984)

\bibitem[M\"{o}bius et al.(2009)]{mobius}M\"{o}bius, E., Bochsler, P.,
Bzowski, M., Crew, G.B., Funsten, H.O., Fuselier, S.A., Ghielmetti, A.,
Heirtzler, D., Izmodenov, V.V., Kubiak, M., Kucharek, H., Lee, M.A.,
Leonard, T., McComas, D.J., Petersen, L., Saul, L., Scheer, J.A.,
Schwadron, N., Witte, M., Wurz, P.: Direct observations of interstellar
H, He, and O by the Interstellar Boundary Explorer. Science
{\bf 326}, 969--971 (2009)

\bibitem[Murray and Dermott(1999)]{mude}Murray, C.D., Dermott, S.F.:
Solar System Dynamics. Cambridge University Press, Cambridge (1999)

\bibitem[Parker(1958)]{parker}Parker, E.N.: Dynamics of
the interplanetary gas and magnetic fields. Astrophys. J.
{\bf 128}, 664--676 (1958)

\bibitem[P\'{a}stor(2009)]{relation}P\'{a}stor, P.: Relation between
various formulations of perturbation equations of celestial mechanics.
arXiv: astro-ph/0907.4005 (2009)

\bibitem[P\'{a}stor et al.(2011)]{flow}P\'{a}stor, P., Kla\v{c}ka, J.,
K\'{o}mar, L.: Orbital evolution under the action of fast interstellar
gas flow. Mon. Not. R. Astron. Soc. {\bf 415}, 2637--2651 (2011)

\bibitem[Poynting(1903)]{poynting}Poynting, J.H.: Radiation in
the Solar System: its effect on temperature and its pressure on small bodies.
Philos. T. R. Soc. Lond. {\bf 202}, 525--552 (1903)

\bibitem[Reach et al.(1995)]{reach}Reach, W.T., Franz, B.A., Welland, J.L.,
Hauser, M.G., Kelsall, T.N., Wright, E.L., Rawley, G., Stemwedel, S.W.,
Splesman, W.J.: Observational confirmation of a circumsolar dust
ring by the COBE satellite. Nature {\bf 374}, 521--523 (1995)

\bibitem[Robertson(1937)]{robertson}Robertson, H.P.: Dynamical effects
of radiation in the Solar System. Mon. Not. R. Astron. Soc. {\bf 97},
423--438 (1937)

\bibitem[Scherer(2000)]{scherer}Scherer, K.: Drag forces on
interplanetary dust grains induced by the interstellar neutral gas.
J. Geophys. Res. {\bf 105}, A5, 10329 (2000)

\bibitem[Wyatt and Whipple(1950)]{wywh}Wyatt, S.P., Whipple, F.L.:
The Poynting--Robertson effect on meteor orbits. Astrophys. J.
{\bf 111}, 134--141 (1950)

\end{thebibliography}
\end{document}